\documentclass[twocolumn,english,aps,prl,noshowpacs,superscriptaddress,amssymb,amsfonts]{revtex4-1}
\usepackage[T1]{fontenc}
\usepackage[latin9]{inputenc}
\usepackage{amsmath}
\usepackage{epstopdf}
\usepackage{graphicx}
\usepackage{amssymb}
\usepackage{color}
\usepackage{bm}
\usepackage{mathrsfs}
\usepackage[low-sup]{subdepth}
\usepackage{multirow}
\usepackage{amsfonts}
\usepackage[english]{babel}
\usepackage[T1]{fontenc}
\usepackage{times}
\usepackage[colorlinks,bookmarks=true,citecolor=blue,linkcolor=red,urlcolor=blue]{hyperref}
\usepackage[tight, FIGTOPCAP, hang, raggedright, nooneline]{subfigure}
\usepackage{verbatim}
\usepackage{dcolumn}
\usepackage{bbold}
\usepackage{babel}

\makeatletter
\newcommand{\be}{\begin{equation}}
\newcommand{\ee}{\end{equation}}
\newcommand{\bea}{\begin{eqnarray}}
\newcommand{\eea}{\end{eqnarray}}

\providecommand{\up}{\uparrow}
\providecommand{\dn}{\downarrow}

\def\avg#1{\left\langle#1\right\rangle}
\newcommand{\ket}[1]{\left|#1\right>}
\newcommand{\bra}[1]{\left<#1\right|}
\newcommand{\braket}[1]{\left<#1\right>}
\newcommand{\para}[1]{\left(#1\right)}
\newcommand{\abs}[1]{\left|#1\right|}

\newcommand{\COMMENT}[1]{}

\def\beq{\begin{equation}}
\def\eeq{\end{equation}}
\def\barray{\begin{eqnarray}}
\def\earray{\end{eqnarray}}
\def\up{\uparrow}
\def\dn{\downarrow}
\def\Eq#1{Eq.~(\ref{#1})}
\def\Fig#1{Fig.~\ref{#1}}

\begin{document}

\title{Observation of Emergent Spacetime Supersymmetry at Superconducting Quantum Criticality}
\author{Zi-Xiang Li}
\affiliation{Department of Physics, University of California, Berkeley, CA 94720, USA}
\affiliation{Institute for Advanced Study, Tsinghua University, Beijing 100084, China}
\author{Abolhassan Vaezi}
\email{vaezi@stanford.edu}
\affiliation{Department of Physics, Stanford University, Stanford, CA 94305, USA}
\author{Christian B. Mendl}
\affiliation{Stanford Institute for Materials and Energy Sciences, SLAC National Accelerator Laboratory and Stanford University, Menlo Park,
California 94025, USA}
\affiliation{Institute of Scientific Computing, Faculty of
Mathematics, Technische Universitat Dresden, 01069 Dresden, Germany}
\author{Hong Yao}
\email{yaohong@tsinghua.edu.cn}
\affiliation{Institute for Advanced Study, Tsinghua University, Beijing 100084, China}
\begin{abstract}
No definitive evidence of spacetime supersymmetry (SUSY) that transmutes fermions into bosons and vice versa has been revealed in nature so far. Moreover, whether spacetime SUSY in 2+1 and higher dimensions can occur or emerge in generic microscopic models remains open. Here, we introduce a lattice realization of a \textit{single} Dirac fermion with attractive Hubbard interactions that preserves both time-reversal and chiral symmetries. By performing numerically-exact sign-problem-free determinant quantum Monte Carlo simulations, we show that the interacting single Dirac fermion in 2+1 dimensions features a superconducting quantum critical point (QCP). More remarkably, we demonstrate that the ${\mathcal N}$=2 spacetime SUSY in 2+1D emerges at the superconducting QCP by showing that the fermions and bosons have \textit{identical} anomalous dimensions 1/3, a hallmark of the emergent SUSY.  To the best of our knowledge, this is the first observation of emergent 2+1D spacetime SUSY in quantum microscopic models. We further show some experimental signatures which can be measured to test such emergent SUSY in candidate systems such as the surface of 3D topological insulators.
\end{abstract}
\maketitle

Spacetime supersymmetry (SUSY) was originally proposed as a fundamental symmetry of nature \cite{Weinberg-book,SUSY_book_Wess, SUSY_book_Nilles, SUSY_book_Haber} more than four decades ago but no experimental evidence of SUSY in particle physics has been confirmed \cite{LHC}. Recently, it has been theoretically argued that SUSY can also spontaneously emerge in certain condensed matter systems \cite{Shenker1984,Balents_susy1998,Fendley_2003a, SS_Lee_2007,Grover_Science_2014, SS_Lee_2014, Yao_PRL_2015a, Yao_PRL_2017a, Yao_2016b, Rahmani_PRL_2015, Huijse_PRL_2015, Bauer_2013a, Yang_2010, Hsieh_PRL_2016a,ZHuang2017arxiv}, e.g., near the superconducting quantum critical point (QCP) of an interacting single-flavored Dirac fermions in 2+1 dimensional systems \cite{Grover_Science_2014, SS_Lee_2014}. However, whether this fascinating $\mathcal{N}$=2 SUSY of a single Dirac fermion can emerge in microscopic lattice models in 2+1 dimensions remains unknown so far.

Dirac fermions are essential ingredients of modern physics that can appear as either elementary particles such as electrons and positrons or emergent quasi-particles, e.g., massless Dirac fermions in graphene~\cite{Graphene,CastroNeto_RMP2012} and on the surface of 3D topological insulators~\cite{Kane_RMP,Qi_RMP}. For a single flavor of massless interacting Dirac fermion in 2+1 dimensions, there are numerous interesting phenomena and theoretical predictions, from emergent spacetime SUSY at the superconducting QCP~\cite{Grover_Science_2014, SS_Lee_2014} to the surface topological order~\cite{Wang_STO_2013, Metlitski_STO_2013, Qi_STO_2013, Fidkowski_STO_2013}, as well as fermion dualities ~\cite{Son_duality}.
Although a single Dirac cone can occur on the surface of three-dimensional topological insulators, studying such interacting problems in {\it microscopic} models in two spatial dimensions have been highly challenging due to the notorious no-go-theorem of fermion-doubling \cite{Nelsen_Ninomiya_1981a}. According to this theorem, it is impossible to realize a single Dirac fermion in local lattice models in two spatial dimensions while respecting time-reversal and chiral symmetries. Usual lattice regularization of a single-flavor Dirac fermion violates some of those symmetry requirements such that existing approaches cannot reveal many fascinating aspects associated with a single Dirac fermion.

Here we introduce a novel two-dimensional lattice model of spin-1/2 fermions that features a single Dirac point at $\Gamma$, with perfectly linear energy dispersion and quantized $\pi$ Berry phase around the $\Gamma$ point, and preserves both time-reversal and chiral symmetries. Fermions in this model can hop along either $x$ or $y$ directions with hopping amplitudes that decay in power-law at long distances. At half-filling, namely when the Fermi level exactly at the neutral point of the single Dirac cone, sufficiently strong attractive interactions between fermions should induce superconductivity in the system. If our lattice regularization can indeed capture low-energy physics of a single Dirac cone, spacetime SUSY could emerge at the superconducting QCP. Consequently, it is highly desired to investigate universal properties of this putative superconducting quantum phase transition by a reliable and non-perturbative method like quantum Monte Carlo (QMC)\cite{BSS,QMC_Cerperly,QMC_Sandvik,PROKOFEV1998,Gull-RMP2011} without encountering the fermion-sign problem \cite{QMC_White,QMC_Troyer,Zaanen2008}. However, QMC methods are sign-problem free only for special classes of interacting models \cite{QMC_Zhang_Wu, Majorana_QMC_Yao_2015, Majorana_QMC_Yao_2016, QMC_Xiang, Berg2012, Huffman2014, Chandrasekharan1999, LeiWang-PRL2015}.

\begin{figure}[t]
\includegraphics[width=5.0cm]{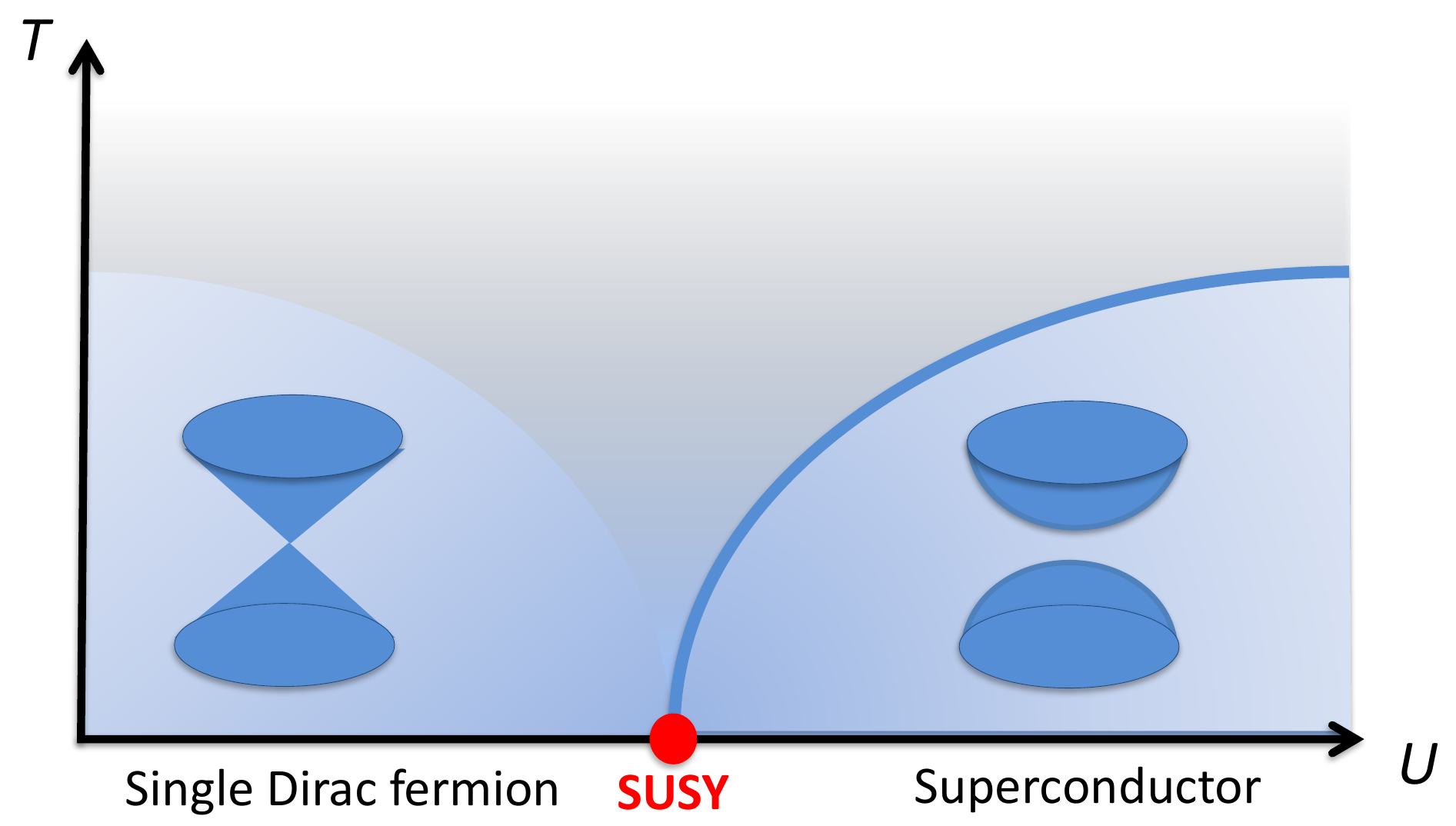}
\caption{The quantum phase diagram of a single-flavor Dirac fermion with onsite attractive interactions. From sign-problem-free start-of-the-art QMC simulations, we show that the $\mathcal{N}\!=\!2$ spacetime SUSY emerges at the superconducting quantum critical point. } \label{fig:phasediagram}
\end{figure}

Remarkably our lattice model of a single Dirac cone with onsite Hubbard attractive interaction $U$ is sign-problem-free, which allows us to study the emergent behaviors of the superconducting quantum phase transition in a numerically {\it exact} way. From the state-of-the-art QMC simulations, we show convincing evidences that the $\mathcal{N}$=2 spacetime SUSY emerges at the superconducting QCP. First, the fermions and order-parameter bosons have \textit{identical} anomalous dimensions that are consistent with the exact value of $1/3$~\cite{Aharony_1997a} associated with the $\mathcal{N}\!=\!2$ SUSY. Moreover, we obtain the correlation-length exponent $\nu\!=\!0.87\!\pm\! 0.05$ which is consistent to the nearly exact result of $0.917$ obtained from conformal bootstrap calculations \cite{Bobev-PRL2015} of the $\mathcal{N}\!=\!2$ SUSY in 2+1 dimensions. Moreover, our QMC calculations show that the local electronic density of states $\rho(\omega)$ at $\omega\ll 1$ behaves like $\rho(\omega)\propto \omega^{a}$ with the exponent $a\!=\!1.37\!\pm\!0.07$, close to the exact value of $4/3$ associated with the $\mathcal{N}\!=\!2$ SUSY, which can be measured in experiments such as STM to test the predicted SUSY. To the best of our knowledge, this is the first numerical observation of emergent spacetime SUSY in 2+1 dimensions.

{\bf The single Dirac fermion model:} To regularize a single Dirac fermion on the square lattice while persevering both time-reversal and chiral symmetries, we introduce the following single-particle Hamiltonian in real space:
\bea
H_0 = \sum_{ij} (t_{ij} c_{i\up}^\dag c_{j\dn} + H.c.),
\eea
where $c^\dag_{i\sigma}$ creates an electron at site ${\bf r}_i$ with spin polarization $\sigma=\up$$/$$\dn$, and $t_{ij}$ is the hopping amplitude between sites ${\bf r}_i$ and ${\bf r}_j$. On the square lattice with $L_x$ ($L_y$) sites along the $x$ ($y$) direction, we are considering the following hoping amplitude $t_{\bf R}$ between two sites separated by ${\bf R}={\bf r}_i-{\bf r}_j$,
\bea
t_{\bf R} = f(R_x,L_x)\delta_{R_y,0} + i f(R_y,L_y)\delta_{R_x,0},
\eea
where $f(R,L)=i \frac{\para{-1}^R}{\frac{L}{\pi}\sin\para{\frac{\pi R}{L}}}$. Note that the feature of hopping only along either $x$ or $y$ directions is not essential and appropriate hopping along other directions can be added without qualitatively changing the main physics discussed below. We now show that the above lattice model satisfies all of the requirements expected for a single Dirac fermion for all practical reasons.

\begin{figure}
\includegraphics[height=2.9cm]{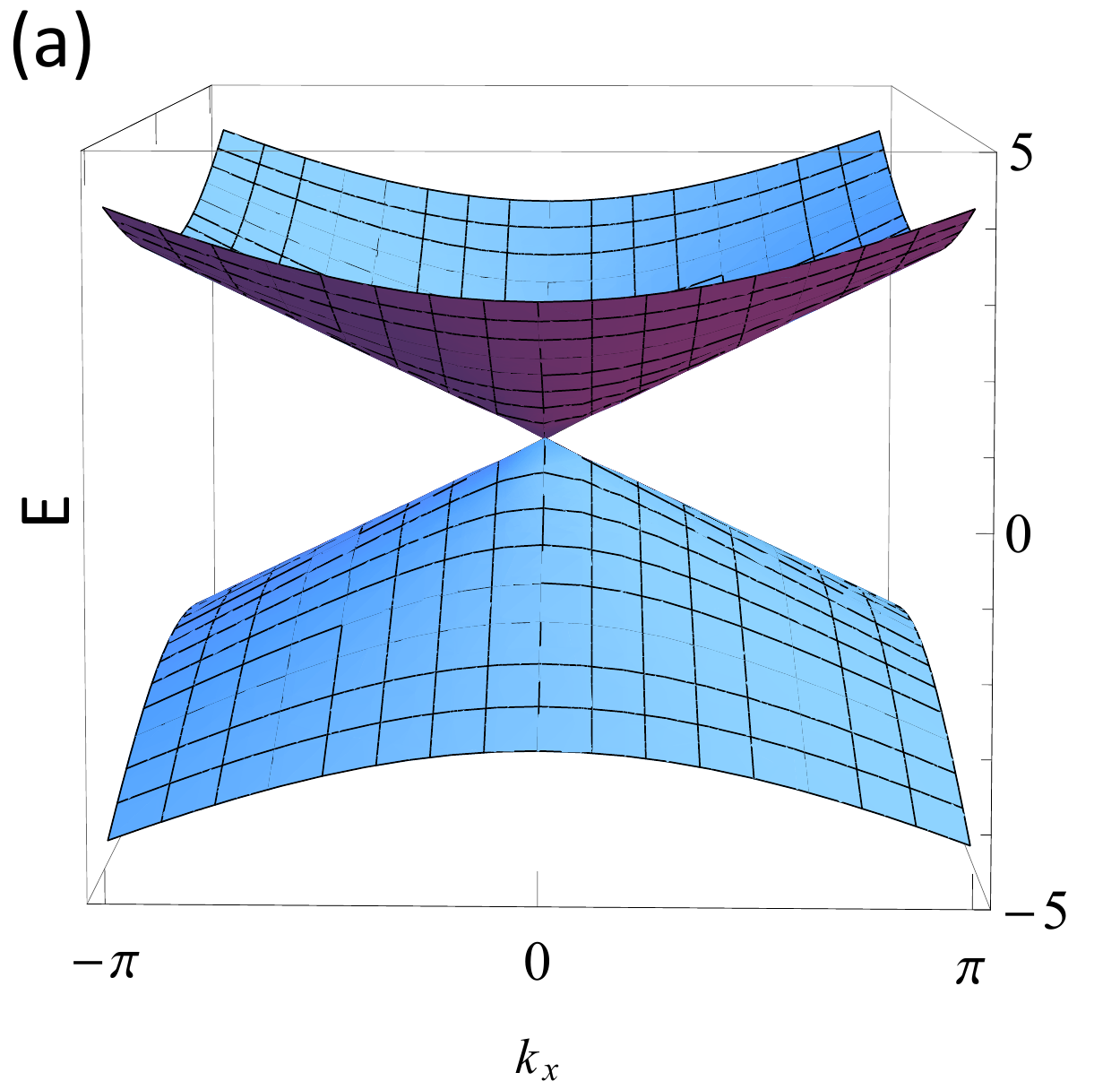}~~~~
\includegraphics[height=2.9cm]{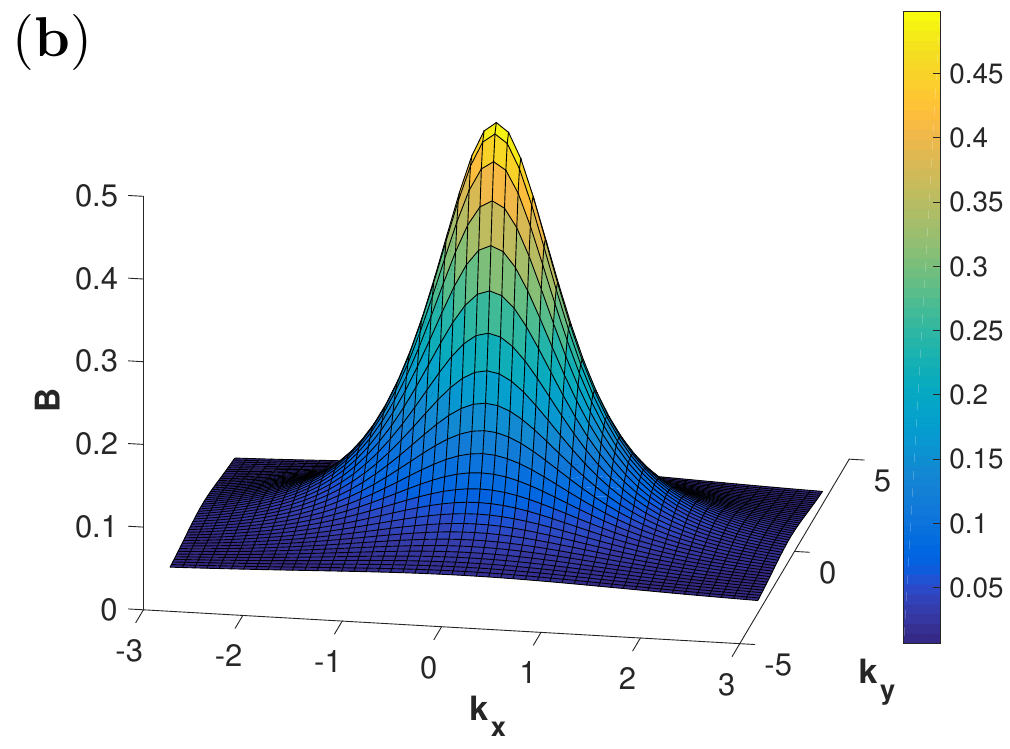}
\includegraphics[height=2.9cm]{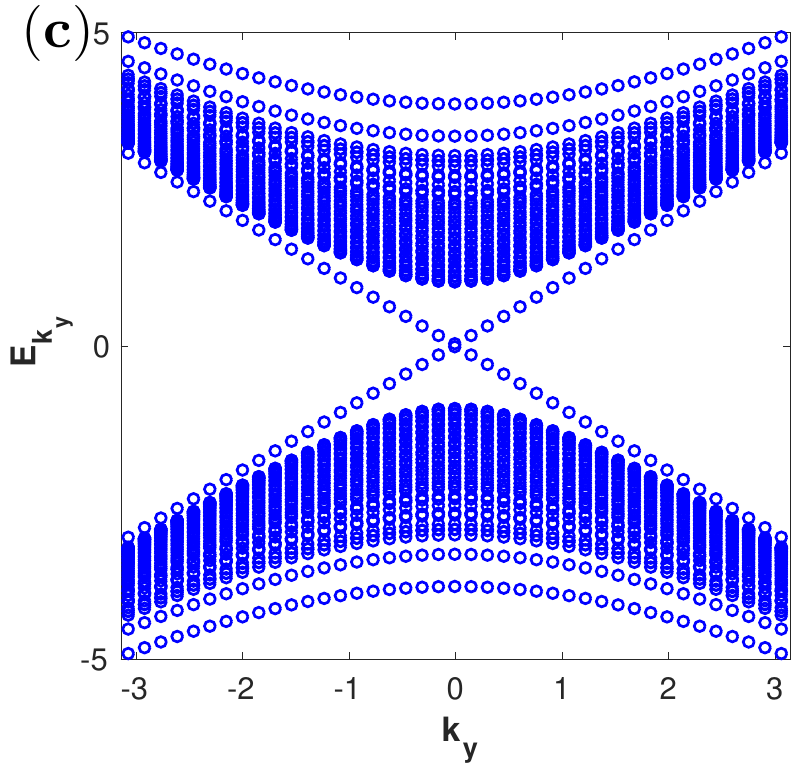}~~~~
\includegraphics[height=2.9cm]{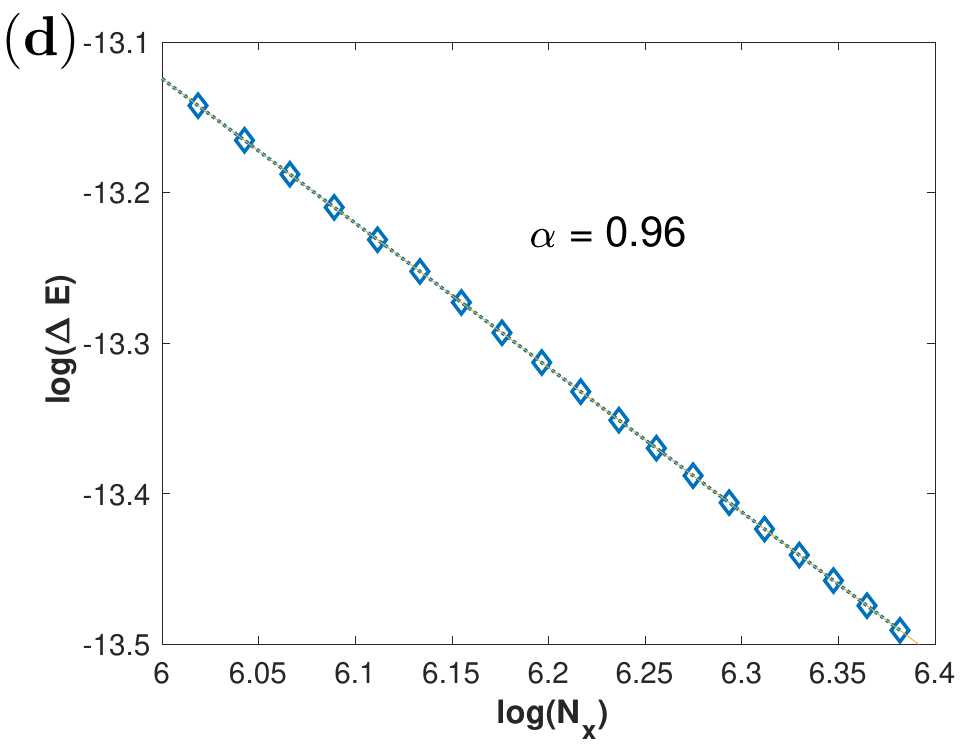}
\caption{The band-structure of lattice fermions with a single Dirac cone. (a) Energy dispersion for a massless lattice Dirac Hamiltonian. The energy dispersion is perfectly linear and given by $E^{\pm}_{k_x,k_y} = \pm \abs{\bf k}$. (b) Berry flux of the same model with a finite mass $m=1$. Although the total Berry phase vanishes, we obtain $\theta_B \simeq \pi$ upon excluding the boundaries of the Brillouin zone. (c) We obtain a nearly gapless left-moving (right-moving) boundary state near $x=0$ ($x=L/2$) by creating a domain wall in the mass term. The results are obtained for $\abs{m} = 1$, and $L_x=L_y=L=41$. The gap at $k_y=0$ is indeed nonzero and equal to $0.0483$. (d) Scaling of the edge-state gap with system size for $\abs{m}=0.1$. This plot implies that $\Delta E \propto 1/L$ when $m\to 0$.} \label{fig:bandstructure}
\end{figure}

It is straightforward to perform Fourier transform to momentum space and obtain
\bea
H_0= \sum_{\bf p} \psi_{\bf p}^\dag v_F\para{p_x\sigma_x+p_y\sigma_y}\psi_{\bf p},
\eea
where $\psi_{\bf p} = \para{c_{{\bf p}\up},c_{{\bf p}\dn}}^{T}$ with $c_{{\bf p}\sigma}$ annihilating a fermion with momentum ${\bf p}=(p_x,p_y)$ and spin $\sigma$, $v_F$ is the Fermi velocity (we set $v_F=1$ from now on), and $\sigma_a$'s denotes Pauli matrices. Note that the momentum eigenvalues ${\bf p}$ run over the first Brillouin zone, and are quantized as $p_{\alpha} = \frac{2n\pi}{L_\alpha}$ for the twisted boundary conditions. It is clear that the lattice model has a single Dirac point at $\bf p$$=$$\bf 0$ (namely $\Gamma$ point) with a linear dispersion all the way to the edge of the first Brillouin zone, as shown in Fig. \ref{fig:bandstructure}(a). Moreover, it can be easily verified that the model is invariant under both time-reversal and chiral symmetries. Note that our model does not directly contradict with the fermion-doubling theorem because the hopping here is not local. In fact, the hopping amplitudes decay as $1/r$ at long distance.

Besides linear dispersion around the single Dirac point, the lattice model above also exhibits most of other physical properties expected for Dirac fermions such as $\pi$ Berry phase around the Dirac point and chiral edge states along mass domain walls. By considering the mass term in the lattice model, namely $H_0\to H_0 +m\sum_i c^\dag_{i}\sigma^z c_i$, it is straightforward to verify that the lattice model gives a Berry phase which is ${\rm sgn}(m)\pi$ for the whole Brillouin zone excluding its boundaries, as shown in Fig. \ref{fig:bandstructure}(b). However, the total Berry phase vanishes due to the $-{\rm sgn}(m)\pi$ contribution of the Brillouin zone boundaries. Although this observation seemly implies the absence of protected zero modes and gapless edge states $\abs{m}>0$ according to the Atiyah-Singer's index theorem~\cite{Atiyah_Singer_1963}, we shall show below that the edge states along domain walls are nearly gapless with a tiny gap that vanishes as $1/L$ where $L$ is the distance between two domain walls.

\begin{figure*}[t]
\includegraphics[height=3.1cm]{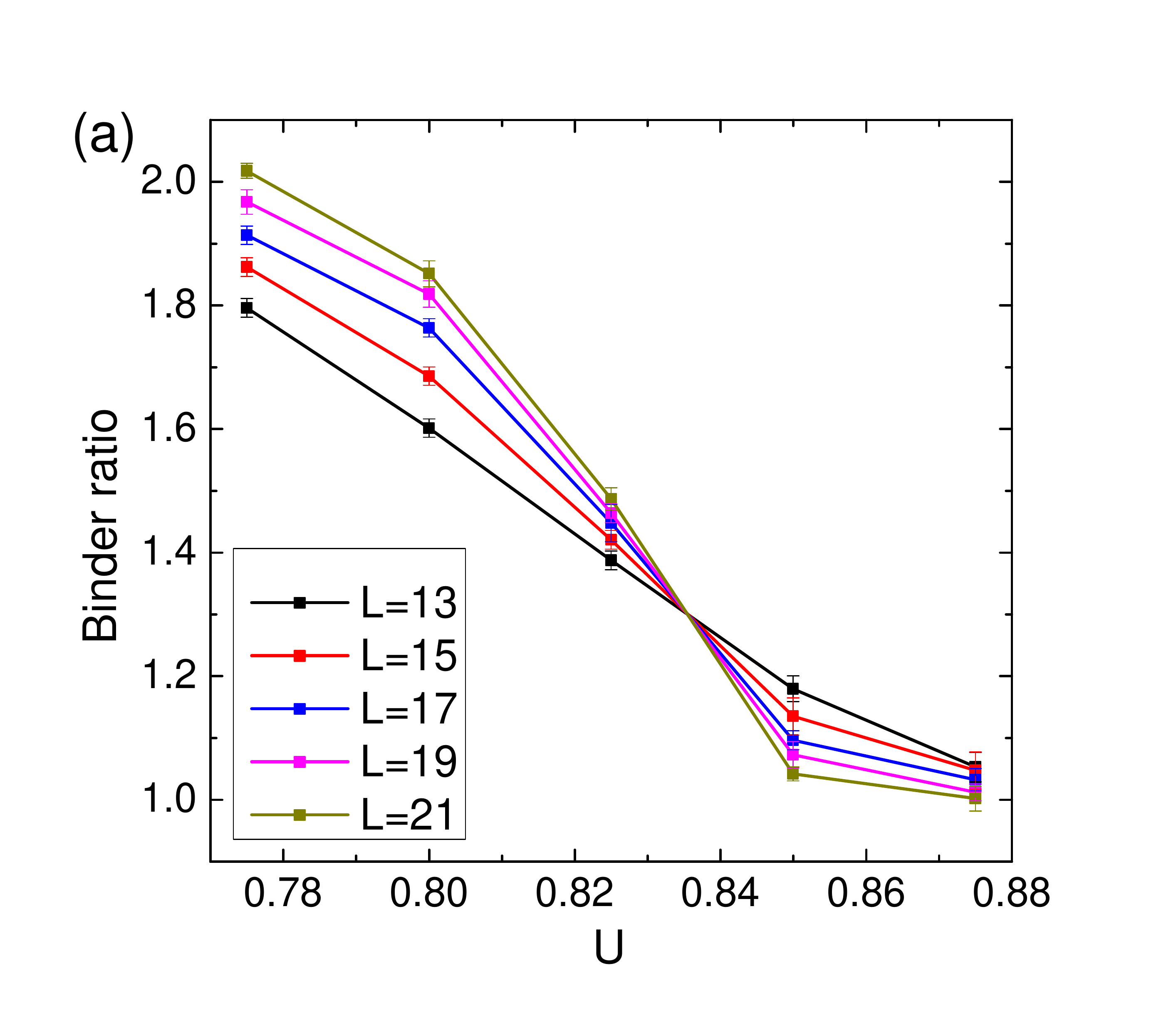}~~~~~~
\includegraphics[height=3.1cm]{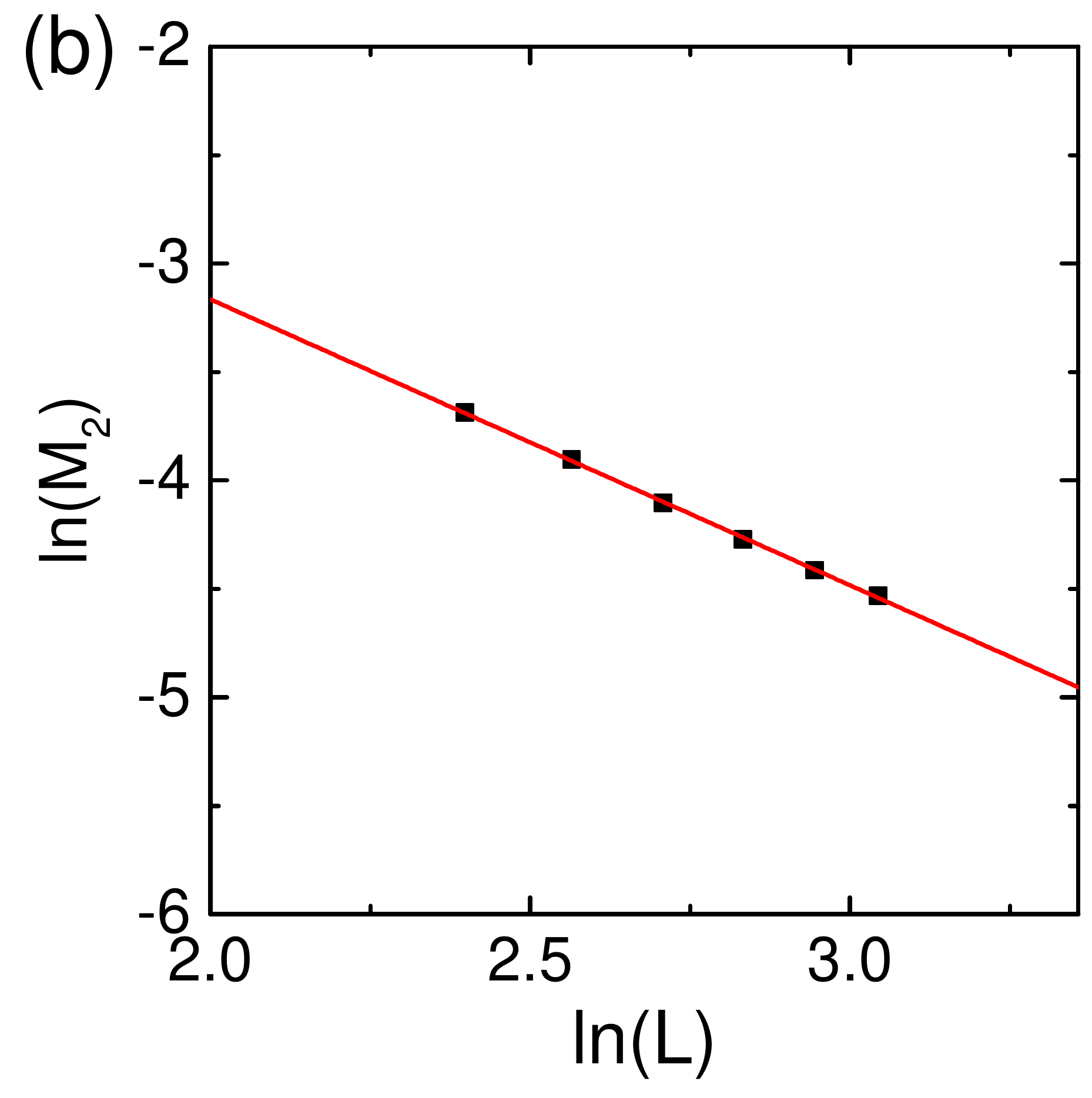}~~~~~~
\includegraphics[height=3.1cm]{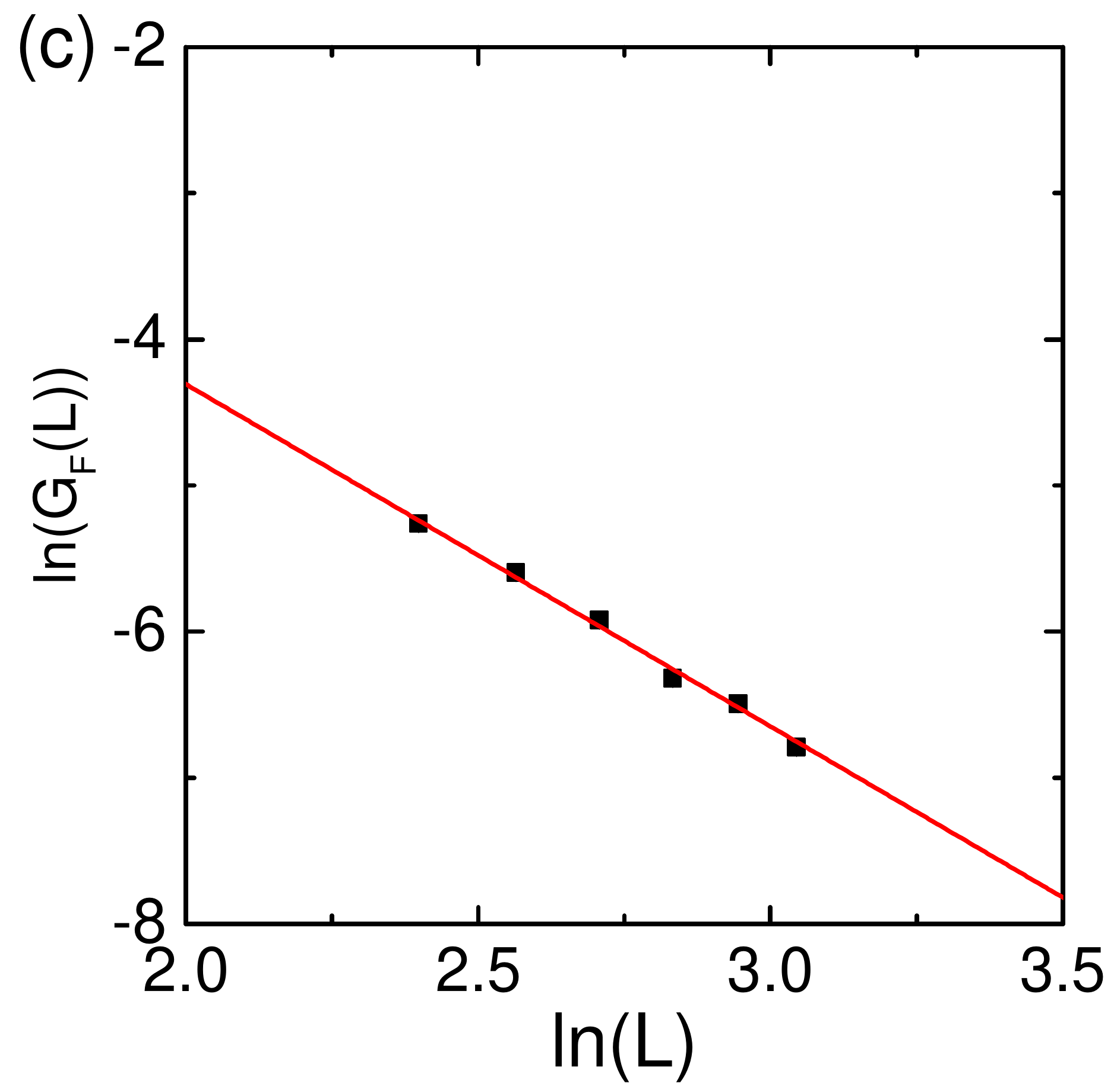}~~~~~~
\includegraphics[height=3.1cm]{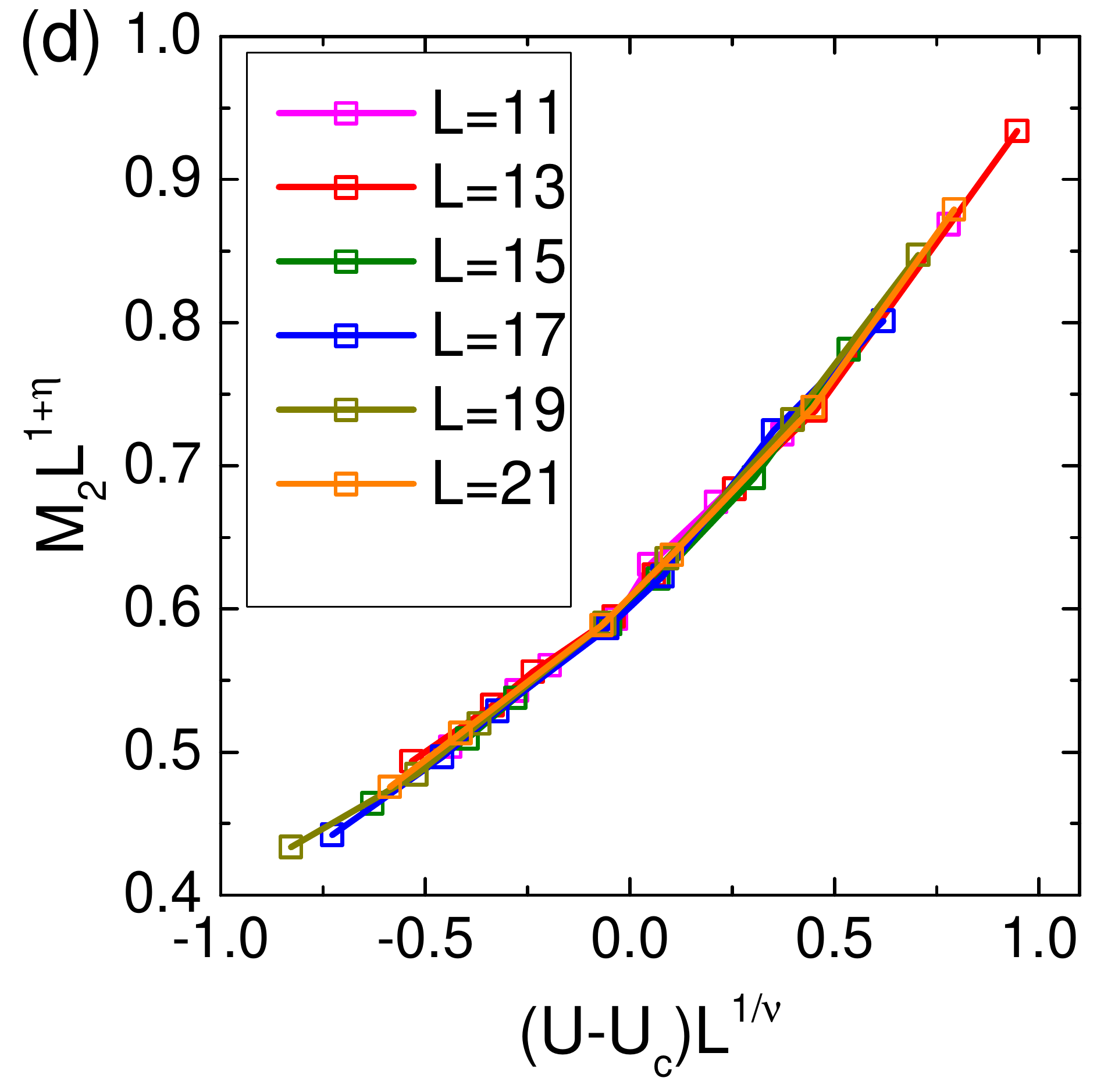}
\caption{The QMC results of superconducting quantum criticality in a single Dirac fermion. (a) The Binder ratio results show that SC phase transition occurs at $U_c \approx 0.83$. (b) From the structure factor of the SC order parameter plotted versus $L$, we obtain the boson anomalous dimension $\eta_b=0.32\pm 0.02$. (c) From the fermion correlation function at largest separation $\vec{R}_{m} \!=\! (\frac{L-1}2\!,\!\frac{L-1}2)$ plotted versus linear system size $L$, we obtain fermion anomalous dimension $\eta_f=0.34\pm 0.05$. (d) Through employing data collapse analysis of SC structure factors near $U=U_c$ for $L=11,\cdots,21$, we obtain the transition point $U_c = 0.827$ and critical exponents $\eta_b = 0.32\pm0.02$, $\nu = 0.87\pm0.05$. } \label{fig:qmc}
\end{figure*}

We now explicitly consider a domain wall for the mass term along $x$ direction and periodic boundary condition along $y$ direction. The local mass term has the following profile: $m\para{x \!<\! L_x/2} = m_0$ and $m\para{x\!\ge\! L_x/2} = -m_0$, where $m_0$ is a finite constant. In Fig. \ref{fig:bandstructure}(c), the energy eigenvalues are plotted against $k_y$. Two nearly gapless modes with opposite chiralities appear, due to the presence of two domain walls. A direct examination of the single-particle wave functions reveals that the chiral (anti-chiral) branch of edge states is localized around $x\!=\!\frac{L_x}{2}$ ($x\!=\!L_x$). For finite $L_x$, due to the nonzero value of the direct hopping between the two domain walls, the edge states exhibit a tiny gap that decays to zero algebraically ($\propto 1/L_x$), as shown in \Fig{fig:bandstructure}(d). The emergence of the chiral modes along boundaries implies that $C^{\rm eff}_{m>0} - C^{\rm eff}_{m<0} = 1$. Recall that $m$$>$$0$ and $m$$<$$0$ regions are time-reversal partners and thus must have opposite Chern numbers, namely $C^{\rm eff}_{m>0} =- C^{\rm eff}_{m<0}$. We thus obtain $C_m^{\rm eff} = {\rm sgn}(m)/2$. Therefore, for all practical reasons, the {\em effective} Chern number of the above hopping model can be considered as ${\rm sgn}(m)/2$, similar to the surface of 3D topological insulators.

{\bf Superconducting quantum criticality:}
Having shown that the regularized lattice model exhibits almost all physical properties of a single Dirac fermion, we are ready to consider interactions in such system with the following Hamiltonian
\bea
H=H_0+\sum_i U(n_{i\uparrow}-1/2)(n_{i\downarrow}-1/2),
\label{interaction}
\eea
where $U$ denotes the strength of onsite Hubbard interactions and $n_{i\sigma}=c^\dag_{i\sigma}c_{i\sigma}$. With the onsite Hubbard interactions, the model still respects the particle-hole symmetry such that the system stays at half-filling. When the Hubbard interaction is attractive, namely $U$$<$$0$, this model is sign-problem free in QMC (the details of QMC are discussed in the Supplementary Materials). Consequently, the interacting effects can be investigated by large-scale numerically exact QMC simulation. Here we employ projector QMC in the Majorana representation to study the ground-state properties as well as nature of quantum phase transitions of the model in Eq. (\ref{interaction}) with attractive Hubbard interaction. (Previously, quantum criticality was only studied by QMC in two-dimensional Dirac semimetals with {\it even} number of Dirac cones, see e.g. Refs. \cite{Assaad-Herbut2013, CDW-Wang2014, CDW-Yao2015, CKXu-PRB2015, Assaad-Herbut-PRB2015, Sorella-PRX2016, Wessel2016,CKXu-PRX2017,charge4e2017,FIQCP2017,Trebst2017SR,HMGuo2017arxiv,Ashvin2017NP,Assaad2016PRX}).

It is expected that singlet superconducting pairing can be generated when the attractive Hubbard interaction is sufficiently strong. To study the quantum phase transition \cite{Subir-book,Sondhi-RMP} into the putative superconducting phase, we calculate the structure factor of onsite singlet pairing on a system with size $L\!\times\!L$: $S_{SC}(L) = \frac{1}{L^4}\sum_{ij}\langle \Delta^\dag_i \Delta_j\rangle$ where $\Delta_j=c_{j\downarrow}c_{j\uparrow}$. The SC long-range order can be extracted through finite-size scaling $\Delta^2_{SC} = \lim_{L\rightarrow\infty} S_{SC}(L)$. Besides, we also measure the quasi-particle excitation gap from time-dependent Green's function. From the state-of-the-art QMC simulations (shown in the SM), we show that the SC order parameter is finite and the single particle gap is opened when the Hubbard interaction exceeds a critical value.

To accurately identify the quantum critical point, we evaluate the RG-invariant quantity Binder ratio that is independent on the system sizes at the critical point. The Binder ratio is defined as $B=\frac{M_4}{M_2^2}$, where $M_2 \equiv \frac{1}{N^2}\sum_{ij}\langle \Delta_i^\dag \Delta_j\rangle$ and $M_4 \equiv  \frac{1}{N^4}\sum_{ijkl}\langle \Delta_i^\dag \Delta_j^\dag \Delta_k \Delta_l\rangle$ for a system with $N\!=\!L^2$ sites. The quantum phase transition point is identified as the crossing point of the Binder ratio for different system sizes $L$. The results for the Binder ratio, as shown in \ref{fig:qmc}(a), convincingly demonstrate that there is a quantum phase transition from the Dirac semimetal phase to the superconducting phase occurring at $U=U_c\approx -0.83$ (in unit of the band width). In the superconducting phase, our QMC calculations show evidences of expected Goldstone modes and Higgs bosons.

{\bf Emergent 2+1D spacetime SUSY:}
At the superconducting QCP $U=U_c$, the system features a single Dirac fermion mode as well as a single complex boson (here the complex boson is the superconducting order-parameter fluctuation). It was argued from the perturbative renormalization-group analysis in $4-\epsilon$ spacetime dimensions that a 2+1D $\mathcal{N}=2$ SUSY might emerge by setting $\epsilon=1$ \cite{Grover_Science_2014, SS_Lee_2014, Yao_PRL_2015a}. However, it is not known {\it a priori} that such spacetime SUSY can emerge in a microscopic model at the QCP and non-perturbative methods such as QMC are needed to address this unambiguously.

If the 2+1D $\mathcal{N}\!=\!2$ SUSY indeed emerges at the superconducting QCP, the anomalous dimensions of fermions and bosons at the QCP should be identical and are equal to 1/3, namely $\eta_f = \eta_b = \frac{1}{3}$. The equivalence of fermion and boson anomalous dimensions is a hallmark of SUSY. To verify whether the superconducting QCP in our model features an emergent spacetime SUSY, we study the critical properties of this quantum phase transition systematically through finite-size scaling (FSS) analysis (the details of FSS are shown in the SM). The anomalous dimensions of the boson and fermion can be extracted via the correlation functions $M_2 \propto \frac{1}{L^{1+\eta_b}}$ and $G_f(L) \!=\! \frac{1}{L^2}\sum_i \langle c^\dag_{i} c_{i+\vec{R}_{m}} + h.c\rangle \!\propto\! \frac{1}{L^{2+\eta_f}}$ according to the definition of anomalous dimensions. Here $\vec{R}_{m}\!=\!(\frac{L-1}{2}\!,\!\frac{L-1}{2})$ is the largest separation between two sites in the system.

The bosonic and fermionic correlation functions at the QCP are shown in \ref{fig:qmc}(b) and \ref{fig:qmc}(c), respectively. Remarkably, the anomalous dimensions of the boson and fermion are equal to each other within errorbar: $\eta_b = 0.32\!\pm\!0.02$ and $\eta_f = 0.34\pm\!0.05$. Moreover, the values of the bosonic and fermionic anomalous dimensions obtained from QMC are consistent with the exact result of 1/3 associated with the 2+1D $\mathcal{N}\!=\!2$ SUSY. These results provide a convincing evidence that the superconducting QCP in our regularized lattice model features emergent SUSY. The consistency between the anomalous dimension of the bosons in the model and the one in the SUSY theory is further supported by the results of the data collapse analysis, as shown in \ref{fig:qmc}(d). In addition, from the data collapse analysis, we extract the correlation-length critical exponent $\nu \!=\! 0.87 \pm\!0.05$, which is consistent with the nearly exact result of 0.917 obtained from the conformal bootstrap calculation of the $\mathcal{N}\!=\!2$ SUSY theory in 2+1D \cite{Bobev-PRL2015} and with the results from RG calculations \cite{Maciejko_PRB_2016a,Zerf2017PRD}. This again indicates that the superconducting QCP in the interacting quantum model features the emergent spacetime SUSY.

\begin{figure}[t]
\includegraphics[height=3.1cm]{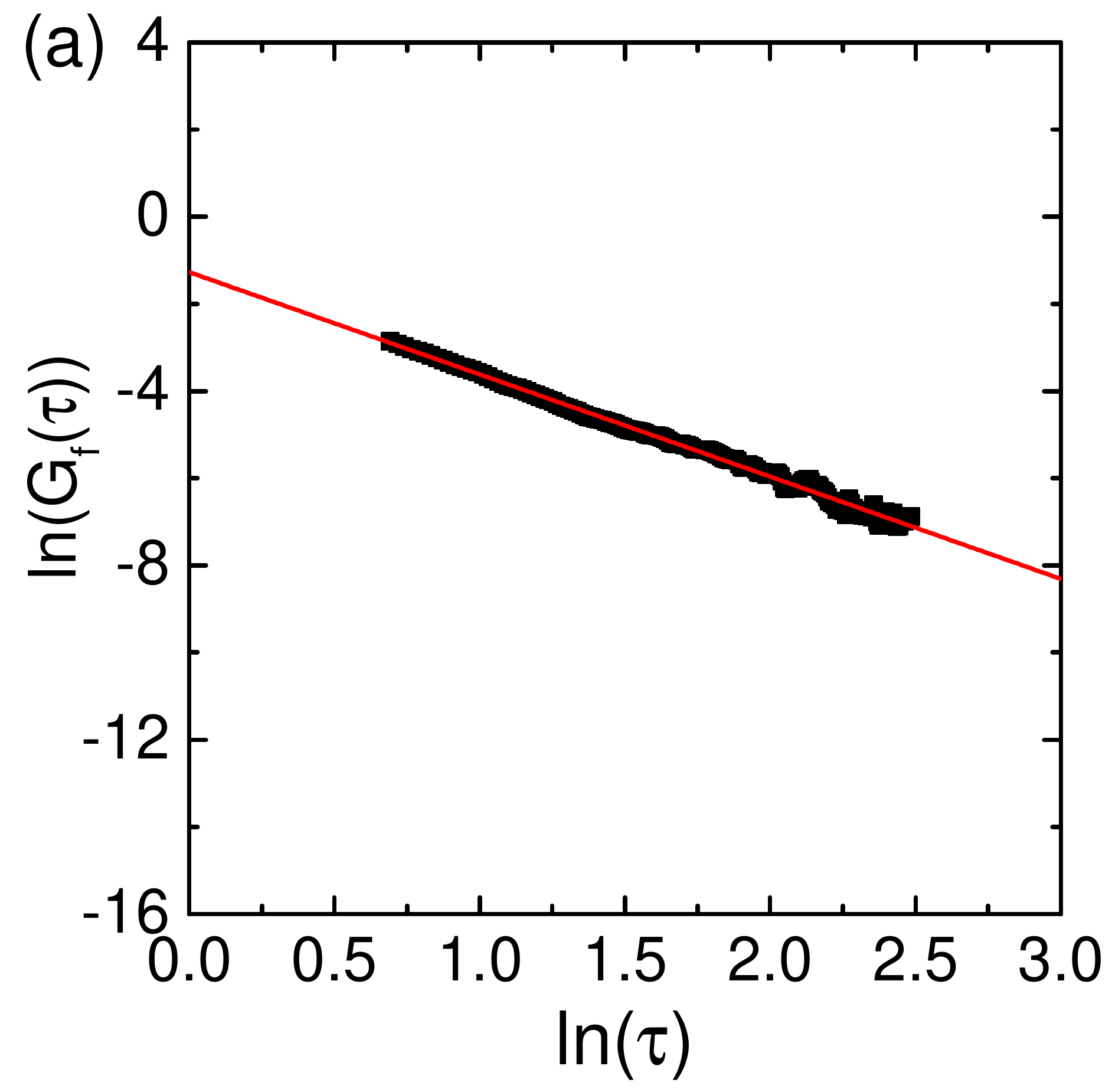}~~~~~~
\includegraphics[height=3.1cm]{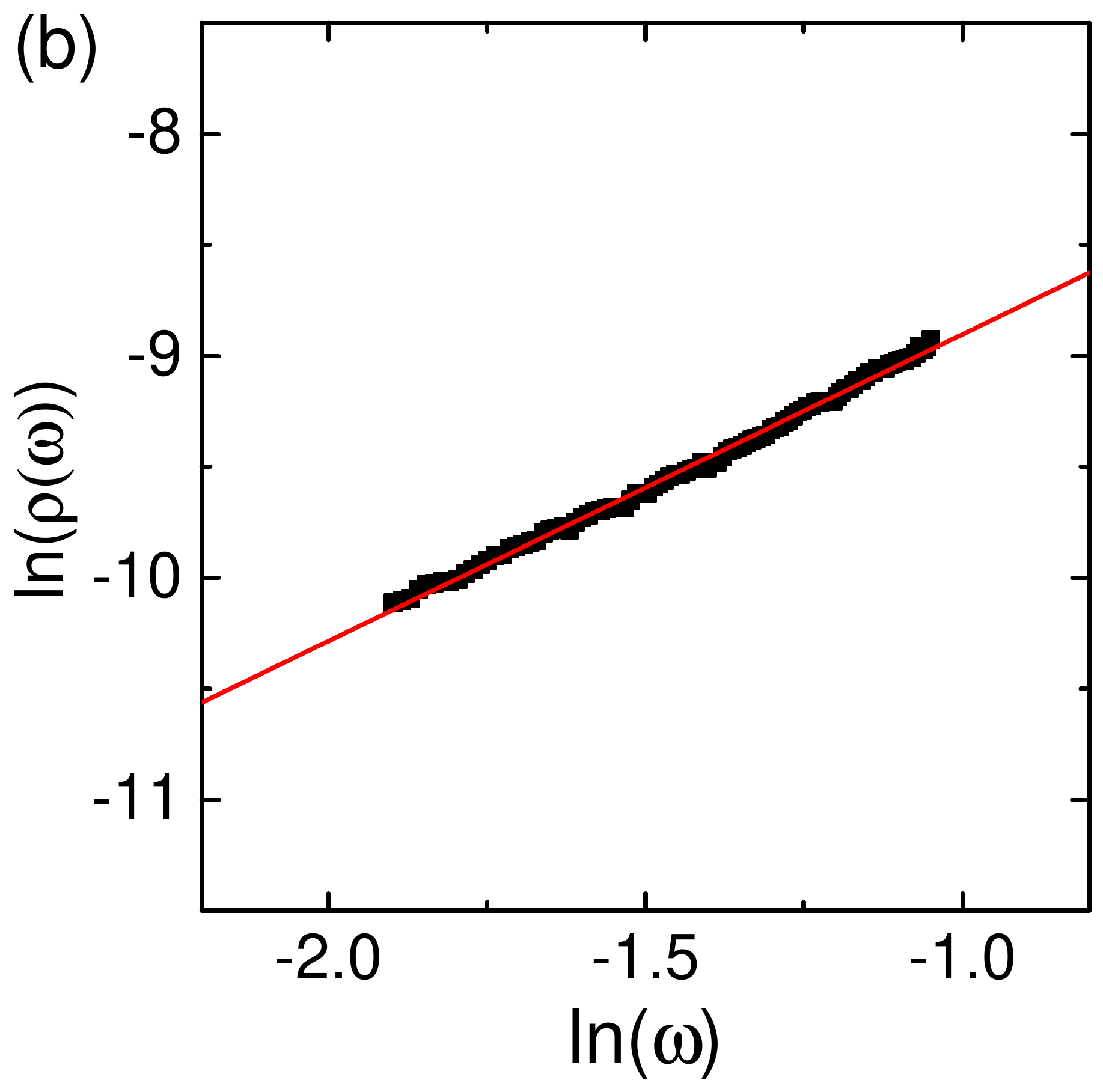}
\caption{The QMC results of LDOS and unequal-time single-particle Green's function at superconducting QCP: (a) From the slope of linear fitting in the ln-ln plot of the Green's function versus imaginary time $\tau$, we obtain $G_f(\tau) \propto \frac{1}{\tau^\alpha}$ with $\alpha = 2.34 \pm 0.02$. (b) From the slope of linear fitting in the ln-ln plot of LDOS versus frequency, we obtain the LDOS $\rho(\omega) \propto \abs{\omega}^a$ with $a= 1.37 \pm 0.07$.  } \label{fig:experiment}
\end{figure}

{\bf Experimental signatures:}
The superconducting QCP of a single Dirac fermion can be potentially observed in realistic materials, such as the surface of 3D topological insulators. There are various experimental ways to check the putative emergent SUSY at the superconducting QCP. For instance, at the QCP the zero-temperature optical conductivity $\sigma(\omega)=K\frac{e^2}{\hbar}$ where $K$ is a constant known exactly due to the emergent SUSY \cite{WWK2016PRL}. Moreover, the SUSY dictates that the local density of states (LDOS) $\rho(\omega)$ of electrons satisfy the scaling law  $\rho(\omega) \propto \abs{\omega}^{\frac{4}{3}}$ for $\omega\ll 1$, which can be measured by STM in experiments. In our model, the LDOS can be calculated by evaluating imaginary-time single-particle Green's function $G_f(\tau) = \avg{c_i(0)c_i(\tau)^\dagger}$ in QMC simulations and then performing analytical continuation (see the SM for details). As shown in \ref{fig:experiment}(a), $G_f(\tau)$ obtained from QMC simulations at the superconducting QCP behaves as $G_f(\tau) \!\propto\! \frac{1}{\tau^\alpha}$ with the exponent $\alpha = 2.34 \pm\!0.02$, which is consistent with the one in the $\mathcal{N}\!=\!2$ SUSY. Moreover, by analytical continuation \cite{Sandvik_1998,Beach_2004}, we obtain the LDOS at superconducting QCP: $\rho(\omega)\propto |\omega|^{a}$ with $a\approx 1.37\pm\!0.07$, as shown in \ref{fig:experiment}(b). This scaling of LDOS is consistent with the exact result of 4/3 given by the $\mathcal{N}\!=\!2$ SUSY within errorbar. More importantly, the LDOS can be measured by STM measurements to experimentally test the emergent SUSY.

{\bf Concluding remarks:}
The emergent SUSY observed at the superconducting QCP in the two-dimensional microscopic model above suggests that the microscopic model can capture all essential physics of a single Dirac cone in 2+1D. In particular, it may be utilized to investigate novel properties of a single Dirac cone on the surface of 3D interacting topological insulators, such as non-Abelian Majorana zero modes at magnetic vortex cones when the single Dirac fermion is superconducting \cite{Fu-Kane_2008}. Moreover, it has been recently argued that it is possible to gap out the single Dirac cone surface states of 3D topological insulators without breaking any symmetry through strong interactions, and the resulting exotic gapped ground state exhibits nontrivial topological order \cite{Wang_STO_2013, Metlitski_STO_2013, Qi_STO_2013, Fidkowski_STO_2013}. One of the approaches to justify the quantum phase transition involves disordering the time-reversal symmetric Fu-Kane state via multiple-vortex proliferation mechanism. We think adding the interaction $V(\sum_{\langle ij\rangle}\Delta_{i}^\dag \Delta_{j}+H.c.)$ with $V$$>$$0$ in \Eq{interaction} can destroy the superconducting phase coherence. It would be interesting to study in the future whether or not such regularized models of a single Dirac fermion can realize the nontrivial surface topological order.

Our work also motivates further studies of other types of 2+1 dimensional spacetime SUSY in microscopic models by non-perturbative methods. For instance, a sign-problem-free microscopic lattice model similar to the one in the present work may be constructed for a single 2+1D massless Majorana fermion that can emerge on the surface of 3+1D topological superconductors. Strong short-range interactions can gap out Majorana fermions by breaking the time-reversal symmetry and its QCP may realize an $\mathcal{N}\!=\!1$ SUSY \cite{Grover_Science_2014}.

Our unbiased and numerically exact simulations of the microscopic quantum model of a single Dirac cone have shown convincing evidence of emergent $\mathcal{N}\!=\!2$ spacetime SUSY in 2+1 dimensions at the superconducting QCP. The results presented here can lend concrete support to potentially realize emergent spacetime SUSY in quantum materials such as the surface of three-dimensional topological insulators, e.g. Bi$_2$Se$_3$. If realized experimentally, it will not only shed light on the intriguing interplay between topology and symmetry, but also provide a promising arena to explore SUSY as well as its spontaneous breaking \cite{Weinberg-book}.

{\it Acknowledgements}: We would like to thank Steve Kivelson for helpful discussions. This work is supported in part by the MOST of China under Grant No. 2016YFA0301001 (H.Y.) and the NSFC under Grant No. 11474175 (Z.-X.L. and H.Y.). A.V. was funded by the Gordon and Betty Moore Foundation's EPiQS Initiative through Grant GBMF4302 and C.B.M. acknowledges support from the DOE Office of Basic Energy Sciences under Grant No. DE-AC02-76SF00515 and from the Alexander von Humboldt Foundation.

%

\begin{widetext}
\section{Supplementary Material}
\renewcommand{\theequation}{S\arabic{equation}}
\setcounter{equation}{0}
\renewcommand{\thefigure}{S\arabic{figure}}
\setcounter{figure}{0}

\subsection{A. Emergent SUSY in interacting 2+1D Dirac model}
In this section, we briefly review the RG analysis of emergent spacetime SUSY at the QCP of a single massless Dirac fermion \cite{Grover_Science_2014, SS_Lee_2014}. To that end, consider a single favor of 2+1D massless Dirac fermions that interact with each other through an attractive Hubbard interaction $U$. The weak Hubbard interaction is irrelevant in 2+1 dimensions, and thus superconductivity can only develop at strong values of $\abs{U}>\abs{U_c}>0$. Near that superconducting QCP at $U=U_c$, the effective action reads
\bea
S & = & \int d^3x \,\,\left[\overline{\psi} \displaystyle{\not}\partial \psi + g\,\left(\phi\,\, \psi^T \sigma^y \psi  + H.c.\right) + |\partial_\tau \phi|^2 + v_b^2|\vec{\nabla} \phi|^2 + r|\phi|^2 + u |\phi|^4\right],  \label{eq:susy3dTI}
\eea
where $\phi$ describes the superconducting order parameter, $v_b$ is the group velocity of bosons, and $\psi = \para{c_{\up},c_{\dn}}^{\rm T}$ denotes Dirac fermions. In the above action we have adopted $\{\gamma_0,\gamma_1,\gamma_2\} = \{\sigma_z,- \sigma_y,\sigma_x \}$ convention. The effective action implies that $\braket{\phi} \neq 0$ which characterizes a SC state takes place for $r < 0$, which in turn generates a finite superconducting gap for Dirac fermions. On the other hand, for $r>0$, we obtain $\braket{\phi} =0$, and as a result Dirac fermions remain massless. It has been recently shown that the QCP of the above effective action ($r=0$) is invariant under the following SUSY transformation generated by $\varepsilon$ Grassman variable provided $u=g^2/2$, and $v_b = v_F$ ($v_F$ is set to 1 implicitly in \Eq{eq:susy3dTI}):
\begin{eqnarray}
\delta \phi = \bar{\varepsilon} \psi ,\quad \delta \psi  = [-\displaystyle{\not}\partial \phi + g \phi^2] \varepsilon
\end{eqnarray}
where $\varepsilon$ is an infinitesimal two-component Grassman variable. Most significantly, it has also been demonstrated that under renormalization group transformations, $u$ and $g^2/2$ flow to the same value at the QCP, so do $v_b$ and $v_F$. Hence, the superconducting QCP of Dirac fermions realizes
an {\it emergent} $\mathcal{N} = 2$ Wess-Zumino SUSY model. Notably, the exact scaling dimensions of fermionic and bosonic fields for $\mathcal{N}=2$ SUSY have been evaluated with $
\Delta_{\phi}= 2/3, \,\Delta_\psi= \Delta_{\phi} +1/2$, corresponding to $
\eta_\phi= \eta_\psi= 1/3 $ anomalous dimensions which is a hallmark of spacetime SUSY.

\subsection{B. Quantum Monte Carlo study of a single Dirac model with Hubbard interaction}
Using the Hubbard-Stratonovic (HS) transformation, it is possible to rewrite the path integral of the Hubbard model in terms of an ensemble of non-interacting fermion path integrals coupled to a local space-time dependent local HS fields. The main steps involve (a) the Suzuki-Trotter expansion of the path integral, namely $\exp\para{-\beta\para{H_K + H_U}}\sim \para{e^{-\beta H_K/N_{\beta}}e^{-\beta H_U/N_{\beta}}}^{N_{\beta}}+O(\beta/{N_{\beta}})^2$ where $\beta = 1/T$, and (b)
the following discrete HS transformations in the density and Cooper channels, respectively:
\bea
&&2e^{-u\para{n_{i,\up}+n_{i,\dn}-1}^2}=\sum_{s=\pm 1}e^{s\lambda_{\rho}\para{n_{i,\up}+n_{i,\dn}-1}}\\
&&2e^{u\para{n_{i,\up}-n_{i,\dn}}^2}=e^{\mu}\sum_{s=\pm 1}e^{s\lambda_{SC}\para{\Delta_{i}+\Delta_{i}^\dag}},
\eea
where, $u = \frac{\beta U}{2N_{\beta}}$, $\Delta_i = c_{i,\dn}c_{i,\up}$, and $\cosh\para{\lambda_{\rho}}=\cosh\para{\lambda_{SC}} = e^{-\mu}$. It is worth mentioning that the last equations have real solutions for $U<0$ only. For each fixed realization of the space-time dependent HS fields, the action can be easily evaluated and equals a fermionic determinant, e.g., the determine of $M_{\rho}\para{\left\{s_{i,\tau}\right\}} = \partial_{\tau}+\beta H_{K}/N+\lambda_{\rho}s_{i,\tau}$ matrix which is $N_xN_yN_{\beta}$ dimensional for the the first choice of the HS transformation. On the other hand,  both of the above two alternative choices for HS fields respect time reversal symmetry, and since the kinetic term associated with massless Dirac fermions is also time-reversal invariant, it can be verified that the eigenvalues of $M_{\rho/SC}$ have Kramer's degeneracy, and thus their determinants are positive definite. Therefore, the QMC of a single Dirac cone for $U<0$ is sign-free. Finally,  Metropolis algorithm is used to sample the configurations with important contributions to the path integral.

We use projector QMC to investigate the ground state properties of the model of single Dirac fermion described by the Hamiltonian in Eq. (1) with attractive Hubbard interaction. In the projector QMC, the expectation value of an observable $O$ in the ground state can be evaluated as:
$\frac{ \bra{\psi_{0}} O \ket{\psi_0}}{ \avg{\psi_{0} \mid \psi_{0} } }   = \lim_{\Theta\rightarrow \infty} \frac{ \bra{\psi_T}e^{-\Theta H } O e^{-\Theta H} \ket{\psi_T}}{ \bra{\psi_T} e^{-2\Theta H} \ket{\psi_T}}$,
where $ \psi_0 $ is the true ground state wave function and $ \ket{\psi_T}$ is a trial wave function which should have a finite overlap with the true ground state wave function. Note that $\Theta$ is projection parameter in the simulation. Although $\Theta\to\infty$ is needed to reach the exact ground state, in numerically calculations a sufficient large $\Theta$ works for practical purposes of obtaining physical quantities with required accuracy. Because of the absence of sign-problem, we can perform large-scale QMC simulations with large system sizes and sufficiently large $\Theta$. In our QMC simulation, we use periodic boundary condition on the square lattice $L\times L$ with largest $L=21$. The imaginary-time projection parameter is $2\Theta = 60/t$ for most systems in the calculation. In the calculation of single particle gap, the systems with large sizes are computed using $2\Theta = 70/t$. We have checked that all the results stay nearly the same when larger $\Theta$ are used, which ensures desired convergence to the limit of $\Theta\to\infty$.

\begin{figure*}
\begin{centering}
\includegraphics[height=4.6cm]{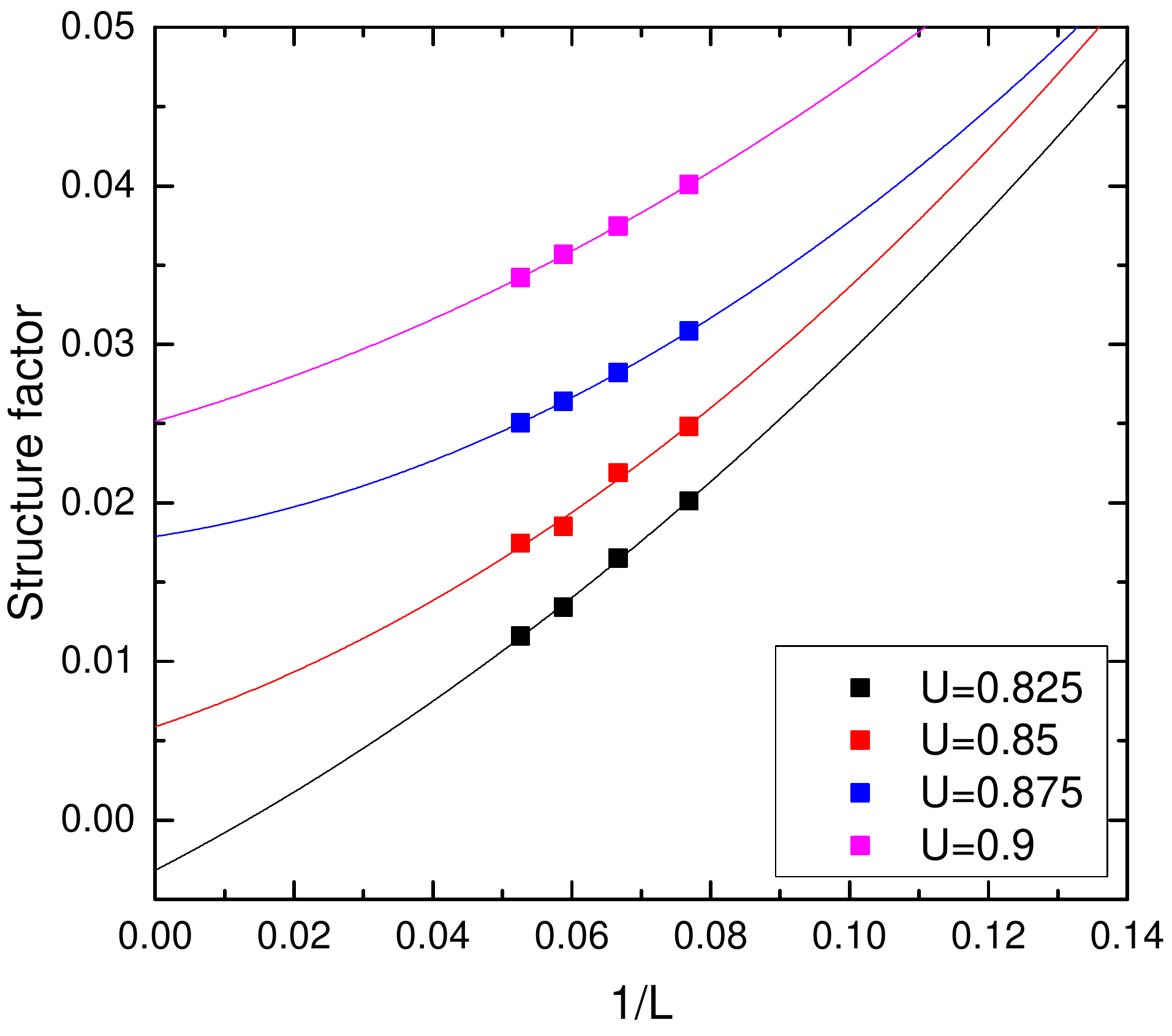}~~~
\includegraphics[height=4.6cm]{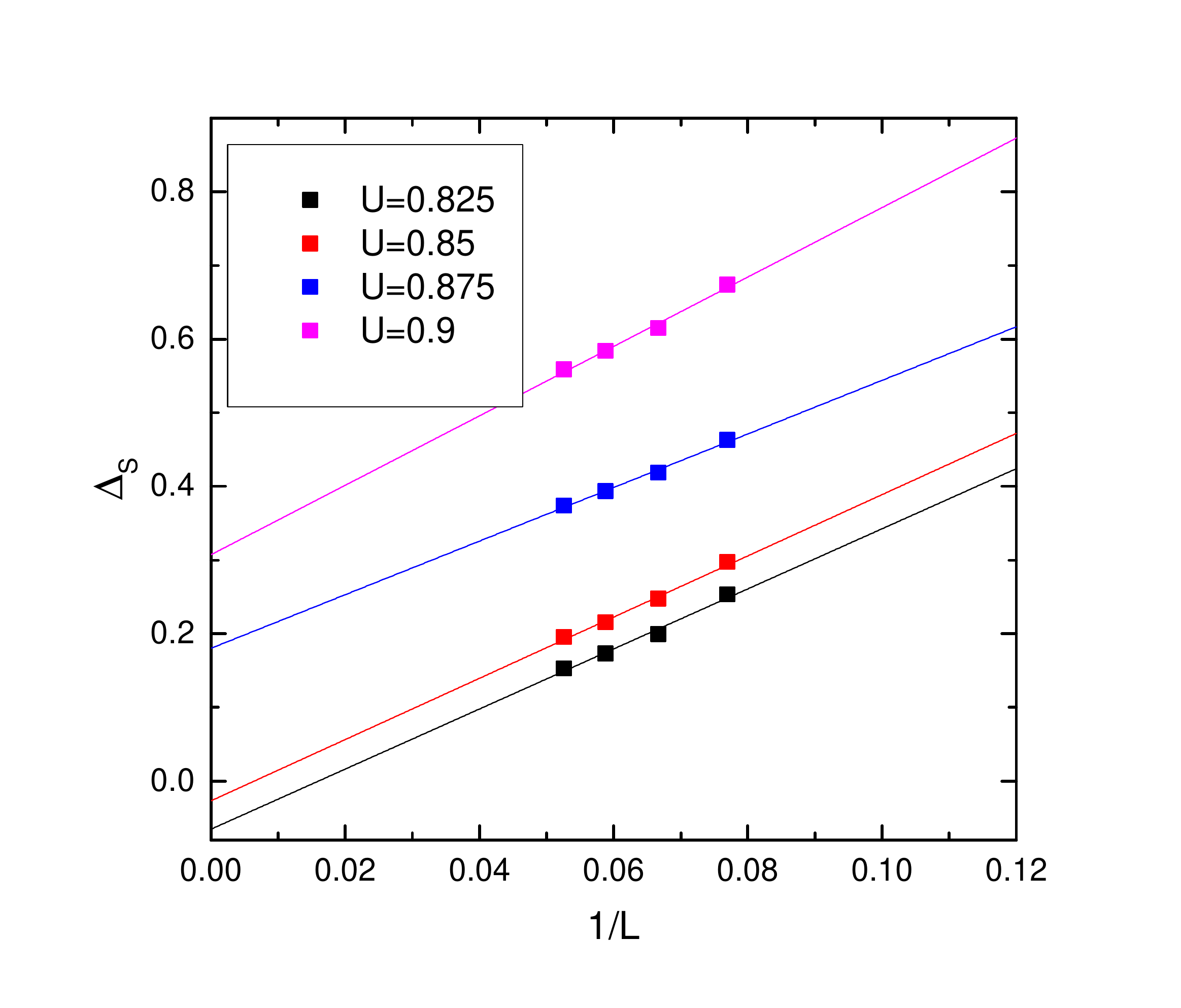}
\par\end{centering}
\caption{The QMC results of SC structure factors and single-particle gaps. (a) Structure factors of SC with $L=13,15,17,19$ for different values of $U$. The data points are fitted by second polynomial curves. (b) Single-particle gap with $L=13,15,17,19$ for different values of $U$. The data points are fitted by linear cures.} \label{fig:qmc2}
\end{figure*}

\subsection{C. Structure factor and single-particle gap}
We employ sign-problem free QMC to compute the structure factors of superconductivity. The results are shown in Fig. \ref{fig:qmc2}. We plot the structure factors for $L=13,15,17,19$ and fit them by second-polynomial curves. The interpolations of the fitted curves are SC structure factors in thermaldynamics limits. From the results of fitting we can see explicitly that the model features SC long-ranged order when $U>0.85$. As the system enters SC phase, it is expected that single-particle gap should be opened by superconducting pairing. In order to verify it, we compute single-particle gap through time-dependent Green's function: $\braket{c_k(0) c_k(\tau)^\dagger} = e^{-\Delta_S(k)\tau}$. We plot the results of single-particle gap for $L=13,15,17,19$ and fit them by linear curves. The results show that single-particle gaps are opened when $U>0.85$.

\subsection{D. Finite size scaling analysis for the superconducting QCP}
To study the critical properties of the superconducting QCP, we perform the finite size scaling analysis. Close to the QCP, the structure factors of SC satisfy the scaling function:
\bea\label{scalingM2}
M_2 = L^{-d-\eta_b + z} {\cal F}( L^{\frac{1}{\nu}}(U-U_c)),
\eea
In our case, we have assumed the dynamical critical exponent $z=1$ for the quantum phase transition to SC phase. $\cal F$ is an unknown scaling function ansatz. When close to the QCP, the structure factors for different system sizes $L$ and different $U$ should be collapsed to a single scaling function if appropriate critical exponents $\nu$ and $\eta$ are chosen. From our scaling analysis, we obtain the critical point $U_c = 0.827$ and critical exponents $\eta_b = 0.32\pm 0.02 $ and $\nu = 0.87\pm0.05 $. The boson anomalous  dimension is consistent with the result extracted from SC order parameter correlation function.

\subsection{E. Imaginary-time single-particle Green's function and local density of state}

The local density of state(LDOS) can be calculated by evaluating the imaginary-time single-particle Green's function $G_f(\tau) = \avg{c_i(0)c_i(\tau)^\dagger}$. Transforming to imaginary-frequency and then performing Matsubara frequency summation yields the formula:
\bea
G_f(\tau) = \int \frac{d \omega}{2\pi} \frac{ e^{-\tau \omega} \rho(\omega)}{e^{-\beta \omega} + 1}
\label{continuation}
\eea

According to the above the formula, the scaling exponents in $\rho(\omega) \propto |\omega|^{a}$ and $G_f(\tau) \propto \frac{1}{\tau^\alpha}$ theoretically satisfy the relation: $\alpha = a + 1$. In QMC simulation, calculating the single-particle Green's function $G_f(\tau)$ with high accuracy is relatively easy, so it is convenient to obtain the scaling exponent $a$ in the LDOS through extracting $\alpha$ in $G_f(\tau)$. As is shown in the main text, we extract the scaling exponent $\alpha = 2.34\pm0.02$ from single-particle Green's function. Through stochastic analytic continuation approach, the LDOS can also be obtained, as shown in Fig. 4(b) of the main text, from which we extracted the scaling exponent $\alpha=1.37\pm 0.07$, which is consistent with the exact value of 4/3 associated with the $\mathcal{N}\!=\!2$ SUSY within errorbar.

\end{widetext}

\end{document}